\documentclass[a4paper,fleqn,usenatbib,useAMS]{mnras}

\usepackage{graphicx}	
\usepackage{amsmath}	
\usepackage{amssymb}	
\usepackage{multicol}        
\usepackage{bm}		

\newcommand{\psr}{PSR~J2032+4127}
\newcommand{\psrb}{PSR~B1259$-$63}
\newcommand{\mttwo}{MT91~213}
\newcommand{\Chandra}{\textit{Chandra}}
\newcommand{\Swift}{\textit{Swift}}
\newcommand{\Fermi}{\textit{Fermi} LAT}
\newcommand{\EW}{\textrm{EW}}

\usepackage[T1]{fontenc}
\usepackage{ae,aecompl}

\usepackage{txfonts}

\title[Monitoring of PSR~J2032+4127/MT91~213]{Multiwavelength monitoring and X-ray brightening of Be X-ray binary PSR~J2032+4127/MT91~213 on its approach to periastron}

\author[W.~C.~G. Ho et al.]{Wynn C. G. Ho$^{1,2}$\thanks{Email: \href{mailto:wynnho@slac.stanford.edu}{wynnho@slac.stanford.edu}},
C.-Y. Ng$^{3}$, Andrew G. Lyne$^{4}$, Ben W. Stappers$^{4}$,
Malcolm J. Coe$^{2}$,
\newauthor
Jules P. Halpern$^{5}$, Tyrel J. Johnson$^{6}$\thanks{Present address: Naval Research Laboratory, Washington, DC, 20375, USA},
Iain A. Steele$^{7}$
\\
$^{1}$Mathematical Sciences and STAG Research Centre, University of Southampton,
Southampton, SO17 1BJ, UK \\
$^{2}$Physics and Astronomy and STAG Research Centre, University of Southampton,
Southampton, SO17 1BJ, UK \\
$^{3}$Department of Physics, the University of Hong Kong, Hong Kong, China \\
$^{4}$Jodrell Bank Centre for Astrophysics, School of Physics and Astronomy,
University of Manchester, Manchester, M13 9PL, UK \\
$^{5}$Columbia Astrophysics Laboratory, Columbia University, New York, NY,
10027, USA \\
$^{6}$College of Science, George Mason University, Fairfax, VA, 22030, USA \\
$^{7}$Astrophysics Research Institute, Liverpool John Moores University,
Liverpool, L3 5RF, UK
}

\date{Accepted 2016 September 19. Received 2016 September 14; in original form 2016 August 19}

\pubyear{2016}

\begin{document}
\label{firstpage}
\pagerange{\pageref{firstpage}--\pageref{lastpage}}
\maketitle

\begin{abstract}
The radio and gamma-ray pulsar \psr\ was recently found to be in a
decades-long orbit with the Be star \mttwo, with the pulsar moving
rapidly towards periastron.
This binary shares many similar characteristics with the previously
unique binary system \psrb/LS~2883.
Here, we describe radio, X-ray, and optical monitoring of \psr/\mttwo.
Our extended orbital phase coverage in radio, supplemented with \Fermi\
gamma-ray data, allows us to update and refine the orbital period to
45--50~yr and time of periastron passage to 2017 November.
We analyse archival and recent \Chandra\ and \Swift\ observations and show
that \psr/\mttwo\ is now brighter in X-rays by a factor of $\sim 70$ since 2002
and $\sim 20$ since 2010.
While the pulsar is still far from periastron, this increase in X-rays is
possibly due to collisions between pulsar and Be star winds.
Optical observations of the H$\alpha$ emission line of the Be star suggest
that the size of its circumstellar disc may be varying by $\sim 2$ over
timescales as short as 1--2~months.
Multiwavelength monitoring of \psr/\mttwo\ will continue through periastron
passage, and the system should present an interesting test case
and comparison to \psrb/LS~2883.
\end{abstract}


\section{Introduction} \label{sec:intro}

Recent radio observations of the 143~ms pulsar \psr\ find it to be part of a
very eccentric,
long orbital period binary system, with the pulsar expected to reach periastron
in late 2017 with its high-mass, Be star companion \citep{lyneetal15}.
The pulsar was discovered by \Fermi\ \citep{abdoetal09} and is
associated with TeV source TeV~J2032+4130 \citep{camiloetal09}.
These characteristics make the \psr\ binary system very similar to the
previously unique pulsar system \psrb/LS~2883 (see, e.g. \citealt{dubus13},
for review), and X-ray results reported here support this similarity.

Subsequent to its discovery in gamma rays, \psr\ was detected in radio by
\citet{camiloetal09}, who also reanalysed a 49~ks \Chandra\ observation taken
in 2004 of the field of the Cygnus OB2 association
(at a distance $d=1.33\pm0.06\mbox{ kpc}$; \citealt{kiminkietal15}),
to which the pulsar likely belongs.
They confirm an X-ray source at the position of the radio source, which also
corresponds to optical source 213 of \citet{masseythompson91}, a B0~V star
(hereafter \mttwo).
\mttwo\ has a mass of either $14.5\,M_{\sun}$ \citep{wrightetal15} or
$17.5\,M_{\sun}$ \citep{kiminkietal07}
and bolometric luminosity
$L_{\rm bol}=1.51\times 10^4\,L_{\sun}=5.79\times 10^{37}\mbox{ erg s$^{-1}$}$
\citep{wrightetal15},
in broad agreement with values appropriate to its spectral and luminosity
classification, i.e.
$15.0\pm2.8\,M_{\sun}$ and $16100\pm130\,L_{\sun}$ \citep{hohleetal10}.
An optical spectrum shows \mttwo\ has a H$\alpha$ equivalent width (\EW) of
$-12.6$~\AA\ \citep{camiloetal09}, which is typical of Be stars and due to
a circumstellar disc surrounding the star.
\citet{camiloetal09} fit a power-law to the \Chandra\ spectrum
and find an unabsorbed 0.5--10~keV X-ray luminosity
$L_{\mathrm X}\approx 6\times 10^{30}\mbox{ erg s$^{-1}$}(d/\mbox{1.3 kpc})^2$.
This X-ray luminosity is compatible with that of either Be stars
(see, e.g. \citealt{berghoferetal97,gagneetal11,nazeetal14}) or
pulsars of age $\sim 10^5\mbox{ yr}$
(see, e.g. \citealt{yakovlevpethick04,pageetal06,potekhinetal15}),
where the pulsar age is taken to be its characteristic spin-down age.

\begin{figure}
 \includegraphics[width=0.9\columnwidth,angle=270,trim={0 4.5cm 0 3.8cm},clip]{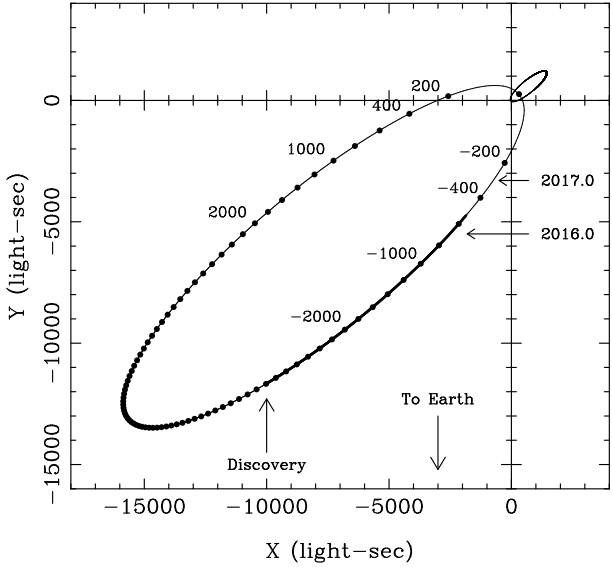}
 \caption{Schematic diagram illustrating the approximate orbital motion of
\psr\ and its Be-star companion \mttwo\ about their common centre of mass.
The orbit shown is that of model~2 (see Table~\ref{tab:spinpars}), which has
an orbital period $P_{\rm b}= 17000$~d and is projected on to the plane
containing the line of sight and major axis of the orbit.
The inclination $i$ of the plane of the orbit to the plane of the sky is
assumed to be 60$^\circ$.
The circles mark 200~d intervals and indicate time from predicted epoch
of periastron, MJD~58069.
The pulsar moves counter-clockwise in the diagram and has been approaching
the Be star since discovery in late 2008/early 2009.
The thick line shows the portion of the pulsar orbit covered by radio
observations reported here, MJD~54689--57538.
Note that orbital velocity is proportional to the separation between the
circles, with a 1000-light-second separation indicating a velocity of about
18~km~s$^{-1}$.
The small ellipse near the origin shows the orbit of the Be star, assuming
that it has a mass of $15\,M_{\sun}$ and that the pulsar has a mass of
$1.35\,M_{\sun}$.}
 \label{fig:orbit}
\end{figure}

More recent analysis of timing observations of \psr\ reveals that its
timing noise can be removed
by considering a timing model in which the pulsar is in an eccentric
(with eccentricity $\epsilon>0.94$), decades-long orbit \citep{lyneetal15}.
Because of the long orbital period, radio measurements up to that time only
cover about 20\% of the orbit, and previous observations (at all wavelengths)
of the pulsar/Be-star binary system have been when the pulsar is on the
apastron-side of the orbit.
Fig.~\ref{fig:orbit} shows a schematic diagram of the system.  Radio and
gamma-ray telescopes continue to monitor the pulsar as it moves towards
periastron.  These observations will refine the orbital parameters, including
eccentricity and mass function, and could, along with VLBI measurements,
directly determine distance and orbital inclination \citep{lyneetal15}.

Many of the above characteristics are typical of Be X-ray binary systems,
albeit with orbital periods of $\lesssim 1\mbox{ yr}$,
which can shine at up to Eddington luminosities
($10^{38}\mbox{ erg s$^{-1}$}$)
when the neutron star nears periastron and accretes matter from the
circumstellar disc
(of size a few times the Be star radius, and larger for isolated Be stars;
\citealt{klusetal14,reigetal16}) of its Be star companion \citep{reig11}.
In this work, we are concerned with X-ray (as well as radio and optical)
emission when the pulsar is far from periastron and not accreting from the
circumstellar disc of the Be star, which might occur near periastron.
Thus the X-ray luminosity is expected to be much lower, as indeed we find,
as well as a brightening that seems in accord with the
well-studied 3.4~yr orbital period gamma-ray binary that contains \psrb.
There are several previous \Chandra\ observations of Cygnus~OB2
(see Table~\ref{tab:obsx}),
and although significantly off-axis, some of these contain the pulsar/Be-star
binary system in the field of view.
As discussed above, \citet{camiloetal09} describe results for the 2004
observation (ObsID~4501), while \citet{rauwetal15} describe results for the
2010 observations (ObsID~10944, 10945, 10951, 10962).

In Section~\ref{sec:radio}, we report results of an updated timing solution
based on monitoring at radio wavelengths, supplemented with \Fermi\ gamma-ray
data.  We (re)analyse all \Chandra\ and \Swift\ X-ray data, as well as our 2016
4.9~ks \Chandra\ Target of Opportunity observation, and report our results in
Section~\ref{sec:xray}.
In Section~\ref{sec:optical}, we report on recent optical measurements
of the H$\alpha$ \EW\ and their implication for the size of
the Be star circumstellar disc.
In Section~\ref{sec:discuss}, we summarize and briefly discuss our findings,
including a few comparisons with \psrb/LS~2883.

\section{Radio and gamma-ray observations and revised timing solution}
\label{sec:radio}

Because of limited orbital phase coverage of observations of the
pulsar since its discovery in 2009, there was strong covariance
between several parameters of the orbital configuration
reported in \citet{lyneetal15}.  Radio timing observations with the 76-m
Lovell Telescope at Jodrell Bank continued during the subsequent two years.
These observations, along with gamma-ray times of arrival (TOAs; see next),
are all processed using the methodology described in \citet{lyneetal15},
which uses the {\sc tempo2} \citep{hobbsetal06} pulsar timing analysis package.
Timing data now span a total of nearly 8 years, i.e.
Modified Julian Date (MJD)~54689--57538.

Data from \Fermi\ are used to provide gamma-ray profiles and additional
TOAs over the timespan of the radio data.
We select Pass 8 LAT data, belonging to the \texttt{SOURCE} class as defined
under the \texttt{P8R2\_SOURCE\_V6} instrument response functions, from the
start of the mission (2008 August 4) to 2016 May 23 (MJD 54682--57527) in a
circular region of interest (ROI) centred on the pulsar position and with a
$15^{\circ}$ radius.  We choose events with energies from 100~MeV to 100~GeV
and require the zenith angle to be $\leq 90^{\circ}$.   We restrict time
intervals to be when LAT was in nominal science operations, data are flagged
as good, and there are no corresponding bright solar flares or gamma-ray bursts.
We perform a binned maximum likelihood analysis on a
$20^{\circ}\times20^{\circ}$ square region, including all sources in the
3FGL catalogue within 25$^{\circ}$ of the ROI centre \citep{aceroetal15}.
Diffuse emission components are modelled using \emph{gll\_iem\_v06.fits}
\citep{aceroetal16} and \emph{iso\_P8R2\_SOURCE\_V6\_v06.txt}
templates\footnote{http://fermi.gsfc.nasa.gov/ssc/data/access/lat/BackgroundModels.html}.
Analysis of spatial and spectral residuals does not reveal the need for any
additional components.

Following the procedure outlined in \citet{abdoetal13}, we model the
gamma-ray spectrum of \psr\ as either a power law, power law with
exponential cutoff, or power law with exponential cutoff and free
exponential index $b$.  We compare the best-fitting likelihood using each model
and find significant curvature ($\sim25\sigma$) and a preference for the
model with $b$ being free ($\sim3\sigma$). The best-fitting model has a power
law photon index of $1.47\pm0.03$, cutoff energy of $4.49\pm0.37$ GeV, and
$b = 0.89\pm0.04$; uncertainties are purely statistical errors.  These
values do not agree, within uncertainties, with those of
\citet{abdoetal13}, but we note that their analysis uses three years of
P7\_V6 data and finds TS$_{b,{\rm free}}=0$ and thus uses $b\equiv1$.
We perform spectral fits in $\sim$30 day bins of 95 flux measurements with
only the normalization
of \psr\ free and calculate a value of TS$_{\rm var}$ (as defined in
\citealt{aceroetal15}) of 109, which is less than the threshold value
of 128.80 and indicates no detectable flux variability
[with a 1--100~GeV flux of
$(2.42\pm0.04)\times 10^{-8}\mbox{ photon cm$^{-2}$ s$^{-1}$}$,
consistent with that of \citealt{aceroetal15}].
Using the best-fit model with $b$ being free and events within $3^{\circ}$
of the ROI centre, we calculate spectral weights to enhance sensitivity
\citep{kerr11}.  We then use the weighted events to construct 95~TOAs
(at $\sim30$-day cadence) as described in \citet{rayetal11}.

\begin{figure}
 \includegraphics[width=\columnwidth,trim={0 1.5cm 0 0},clip]{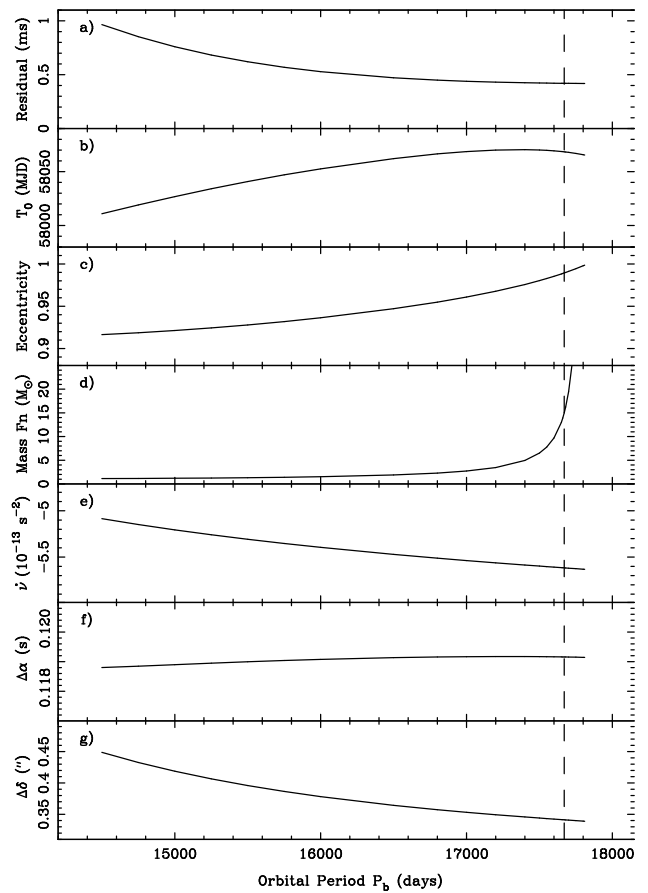}
 \caption{Results of fits to TOAs of a number of binary models having
 different fixed values of orbital period $P_{\rm b}$:
 (a) rms timing residual $\sigma_{\rm t}$,
 (b) epoch of periastron $T_{\rm 0}$,
 (c) orbital eccentricity $\epsilon$,
 (d) mass function $f_{\rm m}$,
 (e) pulsar rotational frequency derivative $\dot\nu$,
 (f) deviation of right ascension of pulsar $\Delta\alpha$ from
$20^{\rm h}32^{\rm m}13^{\rm s}$,
 (g) deviation of declination $\Delta\delta$ from $41^\circ27\arcmin24\arcsec$.
Solutions to the right of the vertical dashed line
(at $P_{\rm b}=17670\mbox{ d}$) are disallowed by the pulsar and
Be star masses ($f_{\rm m}\le 15\,M_{\sun}$; see text).
Parameters for $P_{\rm b}=16000$ (Model 1), 17000 (Model 2), and 17670~d
(Model 3) are given in Table~\ref{tab:spinpars}.
}
 \label{fig:radiofits}
\end{figure}

The newly extended orbital phase coverage reduces covariance between
fitted parameters and allows refinement of the timing model.
While we cannot yet obtain a unique solution because
the orbital period and eccentricity are still highly covariant, there
is a one-dimensional family of {\sc tempo2} fits, which are summarized
in Fig.~\ref{fig:radiofits}, as a function of orbital period.
The rms timing residuals reach an asymptotic minimum at about 0.42~ms for
orbital periods approaching 17830~d, the period at which eccentricity is
close to 1.0 and the system would be marginally bound.
However, the mass function $f_{\rm m}$
[$=(M_{\rm OB}\sin i)^3/(M_{\rm OB}+M_{\rm NS})^2$,
where $M_{\rm NS}$ and $M_{\rm OB}$ are the neutron star and Be star masses
and $i$ is the orbital inclination] has a maximum value of
$\approx 15\,M_{\sun}$
[using $M_{\rm NS}=1.35\,M_{\sun}$, $M_{\rm OB}=17.5\,M_{\sun}$ 
(see Sec.~\ref{sec:intro}), and $i=90^\circ$],
and Fig.~\ref{fig:radiofits} shows that $f_{\rm m}=15\,M_{\sun}$ corresponds
to a binary period of 17670~d.
Inspection of timing residuals from the best-fitting long-period
models indicates that timing residuals are essentially `white', with
little evidence of any systematic departure of TOAs from the models.
For periods shorter than about 16000~d, substantial systematic departures
are seen, resulting in an observed increase in rms residuals.
Any solution with orbital period of $16000<P_{\rm b}<17670$ days, with
corresponding eccentricity of $0.94<\epsilon<0.99$, is likely to be
acceptable.
Parameters of three models spanning this acceptable range are presented in
Table~\ref{tab:spinpars}.
For all such acceptable fits, the epoch of periastron is becoming well
constrained at around MJD~58060(10), i.e. centred on early November 2017.
Figure~\ref{fig:orbit} illustrates one possible configuration of the system for
$P_{\rm b}=17000$~d and $\epsilon=0.96$ (model 2 in Table~\ref{tab:spinpars}).
We note that there is no systematic deviation between radio and
gamma-ray residuals through the whole data set, indicating that the
ephemeris and dispersion measure (DM) are correct and that there is no
detectable variation in DM yet.

\begin{table*}
\caption{Binary model fits to TOAs of \psr.  Fitted values of parameters are
given for three representative models that span the best-fitting range of
orbital periods, i.e. 16000~d (Model 1), 17000~d (Model 2), and 17670~d
(Model 3).  Models with periods shorter than that of model 1 have
unacceptably large timing residuals, while those with periods longer than
that of model 3 are disallowed by the pulsar and Be star masses (see text).
1$\sigma$ uncertainties in the last digit(s) are given in parentheses.
Data used in our analysis spans the range MJD~54689--57538.
The units of time are Barycentric Dynamical Time.}
\label{tab:spinpars}
\begin{tabular}{lccc}
\hline
\multicolumn{1}{c}{Parameter} & \multicolumn{1}{c}{Model 1} &\multicolumn{1}{c}{Model 2} &\multicolumn{1}{c}{Model 3} \\
\hline
Right ascension, $\alpha$ (J2000.0)  & $20^{\rm h}32^{\rm m}13.^{\rm s}119(2)$  & $20^{\rm h}32^{\rm m}13.^{\rm s}119(2)$  & $20^{\rm h}32^{\rm m}13.^{\rm s}119(2)$ \\
Declination, $\delta$ (J2000.0) & $41^\circ27\arcmin24\farcs38(2)$ & $41^\circ27\arcmin24\farcs35(2)$ & $41^\circ27\arcmin24\farcs34(2)$ \\
\\
Epoch of frequency, $t_0$ (MJD)              & 55700.0      & 55700.0      & 55700.0 \\
Frequency, $\nu_0$ (Hz)                                 & 6.980979(5)  & 6.980975(6)  & 6.980973(7) \\
Frequency time derivative, $\dot{\nu}_0$ ($10^{-12}$s$^{-2}$) & $-$0.5396(5) & $-$0.5538(4) & $-$0.5617(5) \\
\\
Orbital period, $P_{\rm b}$ (d)         & 16000     & 17000     & 17670 \\
Epoch of periastron, $T_0$ (MJD)  & 58053(1)  & 58069(1)  & 58068(2) \\
Projected semi-major axis, $x$ (light-second)   & 7138(48)  & 9022(216) & 16335(3737) \\
Eccentricity, $\epsilon$                & 0.936(1)  & 0.961(2)  & 0.989(5) \\
Longitude of periastron, $\omega$ (deg)  & 52(1)     & 40(1)     & 21(5) \\
Mass function, $f_{\rm m}$ ($M_{\sun}$)         & 1.5       & 2.7       & 15.0 \\
\\
Glitch epoch, $T_{\rm g}$ (MJD)                     & 55810.77  & 55810.77  & 55810.77 \\
Frequency, $\Delta\nu_{\rm g}$ ($10^{-6}$ Hz)            & 1.9064(1) & 1.9073(1) & 1.9076(1) \\
Frequency time derivative, $\Delta\dot{\nu}_{\rm g}$ ($10^{-15}$s$^{-2}$)  &  $-$0.501(8)   &  $-$0.545(7)   &  $-$0.564(6)  \\
\\
DM (pc cm$^{-3}$)                                   & 114.68(3) & 114.67(2) & 114.66(2) \\
DM time derivative, DM1 (pc cm$^{-3}$yr$^{-1}$)                & $-$0.02(1)& $-$0.01(1)& $-$0.01(1) \\
\\
rms timing residual, $\sigma_{\rm t}$ (ms)                  & 0.53      & 0.44      & 0.42 \\
\hline
\end{tabular}
\end{table*}

In summary, we conclude that \psr\ is in a very eccentric and
weakly bound system with an orbital period of $\approx$~45--50 years.
We now expect periastron to occur around the beginning of November 2017.
However, we caution that the pulsar is rather young, with a characteristic
age of about 200~kyr, and it may therefore suffer from some rotational
timing noise.
Our conclusions are based on the assumption that the pulsar is a
perfect rotator and is not subject to significant timing noise which
would distort measured Keplerian parameters of the binary system.
On the other hand, the best models show little evidence for any unmodelled
timing noise.
Furthermore, as the pulsar approaches periastron, rapidly-changing Doppler
effects are likely to dominate any effects of timing noise.

\section{X-ray observations and results} \label{sec:xray}

\subsection{\Chandra} \label{sec:chandra}

The field of \psr/\mttwo\ is covered by several archival \Chandra\ imaging
observations made with the ACIS-I detector. We also obtained a targeted
4.9~ks exposure using ACIS-S on 2016 Feb 24.  Details of these observations
are given in Table~\ref{tab:obsx}.

\begin{table*}
 \caption{
\Chandra\ and \Swift\ observations of \psr/\mttwo.
\Chandra\ and \Swift\ count-rates are computed for 0.3--7~keV and 0.3--10~keV,
respectively.
Unabsorbed 0.3--10~keV flux $F_{0.3-10}^{\rm unabs}$
is calculated from count-rate using WebPIMMS and assuming
$N_{\rm H}=7.7\times 10^{21}\mbox{ cm$^{-2}$}$ and a power law $\Gamma=2$.
Errors are $1\sigma$.
}
 \label{tab:obsx}
 \begin{tabular}{lclcccc}
  \hline
 Telescope & ObsID & Date & MJD & Exposure & Count-rate
 & $F_{0.3-10}^{\rm unabs}$ \\
 & & & & ks & ks$^{-1}$ & $10^{-13}$ erg cm$^{-2}$ s$^{-1}$ \\
  \hline
\Chandra & 4358 & 2002 Aug 11 & 52497.83 & 4.9 & $0.8\pm0.4$ & $0.2\pm0.1$ \\
\Chandra & 4501 & 2004 Jul 19 & 53205.09 & 48.7 & $1.6\pm0.2$ & $0.51\pm0.06$ \\
\Swift & 37744001 & 2008 Jun 16 & 54633.04 & 10.8 & $1.0\pm0.4$ & $1.0\pm0.4$ \\
\Chandra & 10944 & 2010 Feb 1 & 55228.46 & 28.3 & $3.6\pm0.3$ & $1.1\pm0.1$ \\
\Chandra & 10945 & 2010 Feb 1 & 55228.80 & 27.9 & $2.3\pm0.2$ & $0.72\pm0.08$ \\
\Chandra & 10951 & 2010 Feb 11 & 55238.58 & 29.2 & $2.6\pm0.3$ & $0.83\pm0.09$ \\
\Chandra & 10962 & 2010 Feb 22 & 55249.66 & 29.4 & $1.9\pm0.2$ & $0.60\pm0.08$ \\
\Swift & 32767001 & 2013 Mar 28 & 56379.88 &  2.7 & $0.5\pm0.4$ & $0.5\pm0.5$ \\
\Swift & 32767002 & 2013 Mar 29 & 56380.17 &  3.9 & $2.0\pm1.0$ & $2.0\pm1.0$ \\
\Swift & 32767003 & 2015 Sep 19 & 57284.13 &  9.2 & $4.6\pm0.7$ & $4.6\pm0.7$ \\
\Swift & 34282001 & 2016 Jan 22 & 57409.71 &  7.6 & $4.8\pm0.8$ & $4.9\pm0.9$ \\
\Chandra & 18788 & 2016 Feb 24 & 57442.35 &  4.9 & $25.9\pm2.3$ & $6.6\pm0.6$ \\
\Swift & 34282002 & 2016 Mar 4 & 57451.00 &  5.9 & $6.2\pm1.0$ & $6.3\pm1.1$ \\
\Swift & 34282003 & 2016 Mar 18 & 57465.38 &  3.4 & $5.2\pm1.3$ & $5.3\pm1.3$ \\
\Swift & 34282004 & 2016 Apr 1 & 57479.04 &  3.4 & $3.8\pm1.1$ & $3.8\pm1.1$ \\
\Swift & 34282005 & 2016 Apr 15 & 57493.69 &  3.7 & $6.2\pm1.3$ & $6.3\pm1.3$ \\
\Swift & 34282006 & 2016 May 1 & 57509.82 &  3.5 & $6.6\pm1.4$ & $6.7\pm1.4$ \\
\Swift & 34282007 & 2016 May 4 & 57512.03 &  2.5 & $8.7\pm1.9$ & $8.9\pm2.0$ \\
\Swift & 34282008 & 2016 May 16 & 57524.20 &  3.3 & $8.3\pm1.6$ & $8.4\pm1.6$ \\
\Swift & 34282009 & 2016 May 27 & 57535.33 &  3.3 & $3.2\pm1.0$ & $3.2\pm1.1$ \\
\Swift & 34282010 & 2016 Jun 1 & 57540.02 &  2.1 & $11.2\pm2.3$ & $11.4\pm2.4$ \\
\Swift & 34282011 & 2016 Jun 11 & 57550.17 &  0.80 & $5.9\pm2.8$ & $6.0\pm2.8$ \\
\Swift & 34282012 & 2016 Jun 18 & 57557.54 &  0.92 & $3.8\pm2.1$ & $3.9\pm2.2$ \\
\Swift & 34282014 & 2016 Jul 4 & 57573.83 &  3.1 & $14.0\pm2.2$ & $14.3\pm2.2$ \\
\Swift & 34282016 & 2016 Jul 10 & 57579.16 &  1.8 & $7.8\pm2.1$ & $7.9\pm2.2$ \\
\Swift & 34282017/8 & 2016 Jul 22 & 57591.45 &  6.9 & $12.7\pm1.4$ & $12.9\pm1.4$ \\
\Swift & 34282023 & 2016 Aug 19 & 57619.25 &  5.0 & $6.6\pm1.2$ & $6.7\pm1.2$ \\
\Swift & 34282024 & 2016 Sep 2 & 57633.16 &  3.4 & $11.4\pm1.9$ & $11.6\pm1.9$ \\
  \hline
 \end{tabular}
\end{table*}

\begin{figure}
 \includegraphics[width=\columnwidth]{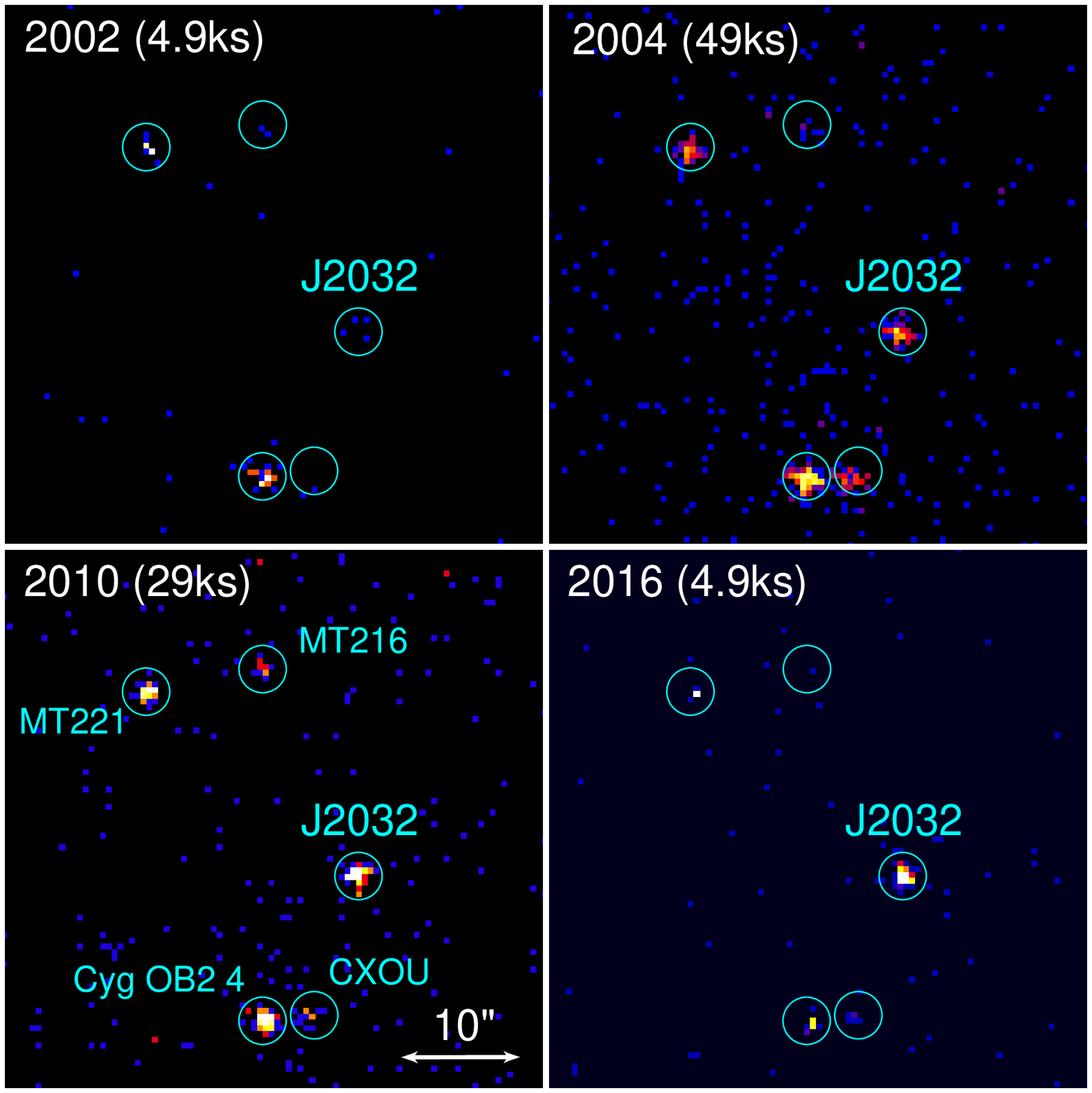}
 \caption{
\Chandra\ field of \psr/\mttwo\ in 2002, 2004, 2010, and 2016, with \psr\ and
four nearby sources (CXOU~J203213.5+412711, Cygnus~OB2~4, MT91~216, and
MT91~221) indicated by circles and labelled.  North is up, and east is left.
}
 \label{fig:field}
\end{figure}

We carry out all \Chandra\ data reduction and analysis using CIAO 4.8 with
CALDB~4.7.1. We reprocess the data using the script \texttt{chandra\_repro},
which applies the latest calibration.  A subset of images in the energy range
0.5--7~keV is shown in Fig.~\ref{fig:field}.
There are four other sources within $\sim$20\arcsec\ of \psr, and their
count-rates are listed in Table~\ref{tab:nearx} (from the 2010 observations,
only ObsID~10951 is used since these sources are located far off-axis in
other exposures).
Cygnus~OB2~4 is a O7~III(f) star \citep{walborn73}.
MT91~216 and MT91~221 are stars 216 and 221, respectively, of
\citet{masseythompson91}, with MT91~216 being a B1.5~V star and MT91~221
being a B2~V star \citep{kiminkietal07}.
From optical spectra, \citet{camiloetal09} identify CXOU~J203213.5+412711
as a Be star that is also in Cygnus~OB2.

\begin{table*}
 \caption{
\Chandra\ observations of sources near \psr/\mttwo.  Count-rates are
computed for 0.3--7~keV.  Errors are $1\sigma$.
}
 \label{tab:nearx}
 \begin{tabular}{lccccc}
  \hline
 Date & \multicolumn{5}{c}{Count-rate (ks$^{-1}$)} \\
  \hline
 & \psr/\mttwo & Cygnus OB2 4 & MT91~221 & CXOU~J203213.5+412711 & MT91~216 \\
2002 Aug 11 & $0.8\pm0.4$ & $5.1\pm1.0$ & $2.2\pm0.7$ & $0.2\pm0.2$
 & $0.4\pm0.3$ \\
2004 Jul 19 & $1.6\pm0.2$ & $4.4\pm0.3$ & $1.2\pm0.2$ & $1.2\pm0.2$
 & $0.1\pm0.1$ \\
2010 Feb 11 & $2.6\pm0.3$ & $3.3\pm0.3$ & $1.8\pm0.2$ & $0.5\pm0.1$
 & $0.5\pm0.1$ \\
2016 Feb 24 & $25.9\pm2.3$ & $3.0\pm0.8$ & $1.8\pm0.6$ & $1.2\pm0.5$
 & $0.2\pm0.2$ \\
  \hline
 \end{tabular}
\end{table*}

\begin{figure}
 \includegraphics[width=\columnwidth]{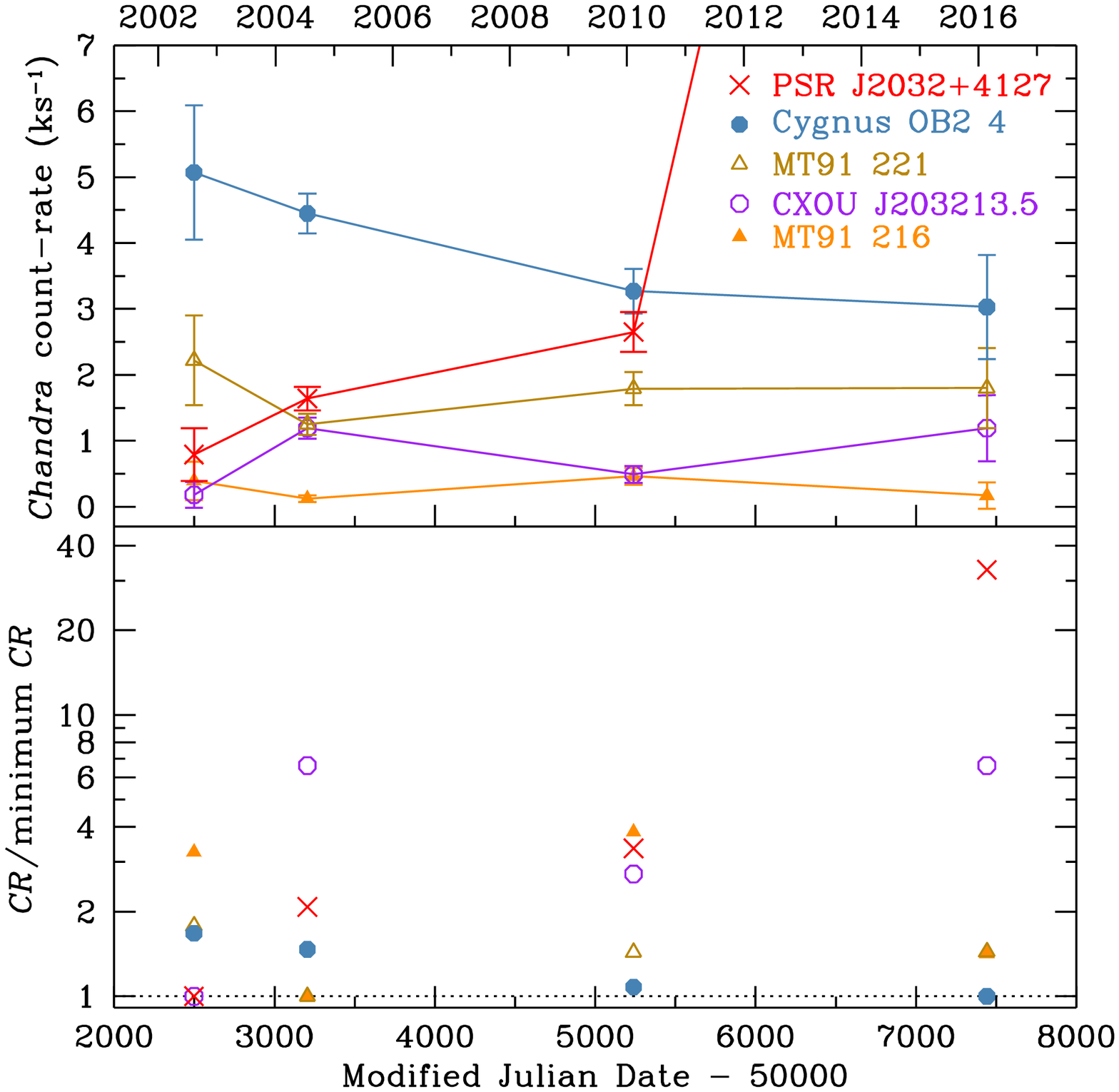}
 \caption{
\Chandra\ light curve of \psr/\mttwo\ and nearby X-ray sources
(see Fig.~\ref{fig:field}).
Top panel: Points (and $1\sigma$ error bars) are 0.3--7~keV
background-subtracted count-rate (see Table~\ref{tab:nearx}).
Note that the 2016 point for \psr\ is at a count-rate of $26\mbox{ ks$^{-1}$}$.
Bottom panel: Points are count-rate ($CR$) from the top panel relative to
minimum count-rate of each source,
i.e. count-rate in 2002 for \psr, 2016 for Cygnus~OB2~4, 2004 for MT91~221,
2002 for CXOU~J203213.5+412711, and 2004 for MT91~216.
}
 \label{fig:nearx}
\end{figure}

Accounting for the different exposure times of each image shown in
Fig.~\ref{fig:field}, it is clear that \psr\ is significantly brighter
at the present time than in previous epochs and relative to neighboring
X-ray sources.
This is demonstrated quantitatively in Fig.~\ref{fig:nearx}, which shows
count-rates as a function of time.
The top panel shows \Chandra\ 0.3--7~keV background-subtracted count-rates,
while the bottom panel shows relative count-rates
and highlights the maximum change in brightness of each source.
\psr\ shows a monotonic increase in brightness, and its variability is
significantly larger (relative increase by a factor of 33) compared to
nearby X-ray sources.
Cygnus OB2 4 and MT91~221, the two brightest sources other than \psr, vary by
$\lesssim 2$.
While the two faint sources CXOU~J203213.5+412711 and MT91~216 seem to vary
by $\lesssim 7$, this variability is possibly due to low counts and
contamination by Cygnus~OB2~4 in the case of CXOU~J203213.5+412711.
Our findings for Cygnus OB2~4, MT91~221, and CXOU~J203213.5+412711 are
consistent with those of \citet{murakamietal11}, who examined variability
between the 2002 and 2004 data.

To perform spectral analyses,
we extract the source spectrum from circular apertures of radius ranging
from 2\farcs5 for on-axis observations to 9\farcs5 for far off-axis
ones. The background spectrum is obtained from nearby source-free regions.
We did not extract a spectrum from the 2002 observation, since only four
photons are detected at the pulsar position.
We assume that the spectrum did not change during the four 2010 observations
and combine these four to improve the signal-to-noise ratio.
Spectral fits are performed using the \emph{Sherpa} package in the 0.5--7~keV
energy range.
Spectra are binned such that there are at least ten counts per bin, and the
statistic from \citet{gehrels86} is used.

\begin{figure}
 \includegraphics[width=\columnwidth]{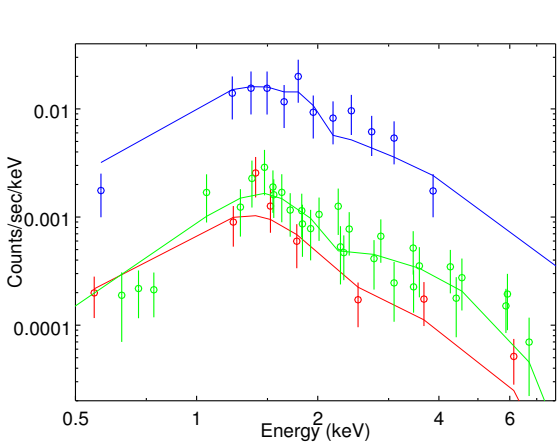}
 \caption{
\Chandra\ ACIS-I spectrum of \psr/\mttwo\ in 2004 (red) and 2010 (green)
and ACIS-S spectrum in 2016 (blue).
Lines are best-fit spectra using an independent power law model for each
observed spectrum and $N_H$ frozen at $7.7\times 10^{21}\mbox{ cm$^{-2}$}$
(see Table~\ref{tab:spectrafit}).
}
 \label{fig:spectrum}
\end{figure}

We fit either a power law model or an optically thin thermal plasma
(\texttt{APEC}) model (see Section~\ref{sec:discuss}) and
accounted for the interstellar absorption using the model \texttt{tbabs}
with \citet{wilmsetal00} abundances.
We first fit all the spectra independently.
We then fit all the spectra with a single absorption column density
$N_{\rm H}$.
Next we assume $N_{\rm H}=7.7\times 10^{21}\mbox{ cm$^{-2}$}$ (based on the
color excess of \mttwo; \citealt{camiloetal09}) and fit each spectrum
independently and jointly
(see Fig.~\ref{fig:spectrum} for results using the power law model).
Parameters between different epochs are formally consistent, although
fit parameters are not well constrained.
Therefore we also fit spectra jointly with the same parameters but allow
different normalization between observations.
All results are listed in Table~\ref{tab:spectrafit}, with $1\sigma$ errors.
Our results using a power law or \texttt{APEC} model for the 2004 data
are consistent with those in \citet{camiloetal09}, who find
$\Gamma=2.1\pm0.7$, $kT=4_{-2}^{+9}\mbox{ keV}$, and
$F_{0.5-10}^{\rm unabs}=0.32\times 10^{-13}$ erg cm$^{-2}$ s$^{-1}$.
For the 2010 data, our results with the \texttt{APEC} model are comparable
to those in \citet{rauwetal15}, who find
$N_{\rm H}=7.9\times 10^{21}\mbox{ cm$^{-2}$}$,
$kT=4_{-1}^{+2}\mbox{ keV}$, normalization~$=2.4\times 10^{-5}$,
and $F^{\rm abs}=0.26\times 10^{-13}$ erg cm$^{-2}$ s$^{-1}$.
Incidentally, a thermal plasma model fit to the spectra of \psrb\ near
apastron gives $kT\approx 6-14\mbox{ keV}$ \citep{hirayamaetal99}.
Finally we fit all observations with the same normalization, but this yielded
poor results ($\chi_\nu^2>2$).
It is important to note that while the models we use generally yield good
fits and parameters that are consistent within the large errors, the
spectra are of relatively low quality, given the low number of counts
($<130$ counts in each observation) in the early epochs when \psr\ is dim
and the short exposure when \psr\ is bright.

\begin{table}
 \caption{
Spectral fit results of \Chandra\ data of \psr/\mttwo.
Absorption column density $N_{\rm H}$ is in units of $10^{21}\mbox{ cm$^{-2}$}$,
$kT$ is in units of keV, normalization is in units of $10^{-5}$,
and absorbed 0.5--7~keV flux $F_{0.5-7}^{\rm abs}$ is in units of
$10^{-13}\mbox{ erg cm$^{-2}$ s$^{-1}$}$.  Errors are $1\sigma$,
and some upper/lower confidence limits are not constrained.
$^\ast$ indicates that the value of $N_{\rm H}$ is frozen.
}
 \label{tab:spectrafit}
 \begin{tabular}{cccccc}
  \hline
Year & $N_{\rm H}$ & $\Gamma/kT$ & Normalization & $F_{0.5-7}^{\rm abs}$
  & $\chi^2/\mbox{dof}$ \\
  \hline
\multicolumn{6}{c}{all parameters free and independent fit of all observations} \\
 \multicolumn{6}{c}{power law} \\
2004 & $1^{+17}$        & $1.8_{-0.6}^{+2.0}$ & $0.4_{-0.1}^{+2}$   & $0.2_{-0.1}^{+0.1}$  & 3.6/3 \\
2010 & $1^{+7}$         & $1.5_{-0.2}^{+0.5}$ & $0.7_{-0.2}^{+0.6}$ & $0.4_{-0.2}^{+0.1}$ & 10.7/24 \\
2016 & $17_{-12}^{+15}$ & $2.7_{-0.9}^{+1.0}$ & $30_{-20}^{+70}$    & $3_{-3}^{+9}$       & 4.0/8 \\
 \multicolumn{6}{c}{\texttt{APEC}} \\
2004 & $10^{+10}$      & $1.5_{-0.6}$        & $2_{-1}^{+2}$       & $0.1_{-0.07}^{+0.07}$ & 4.6/3 \\
2010 & $0^{+6}$        & $16_{-12}$          & $2.9_{-0.4}^{+1.0}$ & $0.4_{-0.15}^{+0.03}$ & 10.9/24 \\
2016 & $14_{-9}^{+10}$ & $2.1_{-0.7}^{+2.3}$ & $50_{-20}^{+30}$    & $2_{-1}^{+1}$      & 3.5/8 \\
\\
\multicolumn{6}{c}{$N_{\rm H}$ tied and joint fit of observations} \\
 \multicolumn{6}{c}{power law} \\
2004 & $4^{+6}$ & $2.1_{-0.8}^{+0.9}$ & $0.6_{-0.3}^{+0.7}$ & $0.2_{-0.2}^{+1}$ & 20.0/37 \\
2010 &      --- & $1.7_{-0.4}^{+0.5}$ & $1.0_{-0.4}^{+0.7}$ & $0.4_{-0.3}^{+0.6}$ & --- \\
2016 &      --- & $1.8_{-0.5}^{+0.6}$ & $8_{-4}^{+7}$       & $3_{-3}^{+6}$  & --- \\
 \multicolumn{6}{c}{\texttt{APEC}} \\
2004 & $3^{+6}$ & $4_{-3}$       & $1.4_{-0.4}^{+0.7}$ & $0.16_{-0.07}^{+0.06}$ & 20.3/37 \\
2010 &      --- & $8_{-4}^{+76}$ & $3.0_{-0.5}^{+1.0}$ & $0.4_{-0.1}^{+0.1}$ & --- \\
2016 &      --- & $5_{-3}^{+32}$ & $24_{-5}^{+11}$     & $3_{-2}^{+2}$  & --- \\
\\
\multicolumn{6}{c}{$N_{\rm H}$ frozen and independent fit of all observations} \\
 \multicolumn{6}{c}{power law} \\
2004 & $7.7^\ast$ & $2.5_{-0.6}^{+0.7}$ & $0.9_{-0.3}^{+0.4}$ & $0.14_{-0.07}^{+0.1}$ & 3.8/4 \\
2010 & --- & $1.9_{-0.2}^{+0.3}$ & $1.3_{-0.3}^{+0.3}$ & $0.4_{-0.2}^{+0.3}$ & 11.8/25 \\
2016 & --- & $2.1_{-0.3}^{+0.3}$ & $11_{-3}^{+3}$      & $3_{-1}^{+1}$  & 4.7/9 \\
 \multicolumn{6}{c}{\texttt{APEC}} \\
2004 & $7.7^\ast$ & $2.1_{-0.8}^{+2.9}$ & $1.9_{-0.4}^{+0.4}$ & $0.12_{-0.07}^{+0.05}$ & 4.5/4 \\
2010 & --- & $4_{-1}^{+3}$       & $3.8_{-0.5}^{+0.4}$ & $0.36_{-0.1}^{+0.08}$ & 12.4/25 \\
2016 & --- & $3.1_{-0.9}^{+2.1}$ & $32_{-5}^{+5}$      & $2.6_{-0.7}^{+0.5}$  & 4.1/9 \\
\\
\multicolumn{6}{c}{$N_{\rm H}$ and $\Gamma/kT$ tied and joint fit of observations} \\
 \multicolumn{6}{c}{power law} \\
2004 & $3^{+6}$ & $1.7_{-0.4}^{+0.4}$ & $0.4_{-0.2}^{+0.3}$ & $0.2_{-0.06}^{+0.1}$ & 20.4/39 \\
2010 &      --- &                 --- & $0.9_{-0.3}^{+0.7}$ & $0.4_{-0.1}^{+0.3}$ & --- \\
2016 &      --- &                 --- & $7_{-3}^{+5}$       & $3_{-1}^{+2}$  & --- \\
 \multicolumn{6}{c}{\texttt{APEC}} \\
2004 & $2^{+5}$  & $7_{-3}^{+22}$ & $1.3_{-0.3}^{+0.4}$ & $0.17_{-0.04}^{+0.06}$ & 20.6/39 \\
2010 &       --- &            --- & $3.0_{-0.5}^{+0.9}$ & $0.4_{-0.07}^{+0.1}$ & --- \\
2016 &       --- &            --- & $24_{-5}^{+8}$      & $3_{-0.6}^{+1}$  & --- \\
\\
\multicolumn{6}{c}{$N_{\rm H}$ frozen and $\Gamma/kT$ tied and joint fit of observations} \\
 \multicolumn{6}{c}{power law} \\
2004 & $7.7^\ast$ & $2.0_{-0.2}^{+0.2}$ & $0.6_{-0.1}^{+0.2}$ & $0.2_{-0.1}^{+0.2}$ & 21.1/40 \\
2010 & --- &                 --- & $1.4_{-0.2}^{+0.3}$ & $0.35_{-0.06}^{+0.06}$ & --- \\
2016 & --- &                 --- & $11_{-2}^{+2}$      & $2.8_{-0.3}^{+0.4}$  & --- \\
 \multicolumn{6}{c}{\texttt{APEC}} \\
2004 & $7.7^\ast$ & $4_{-0.8}^{+1}$ & $1.7_{-0.3}^{+0.3}$ & $0.15_{-0.07}^{+0.04}$ & 21.9/40 \\
2010 & --- &           --- & $3.9_{-0.4}^{+0.5}$ & $0.35_{-0.06}^{+0.06}$ & --- \\
2016 & --- &           --- & $31_{-4}^{+5}$      & $2.8_{-0.4}^{+0.4}$  & --- \\
  \hline
 \end{tabular}
\end{table}

\subsection{\Swift} \label{sec:swift}

To improve our understanding of the X-ray light curve of \psr, we checked
archival data taken by \Swift\ \citep{burrowsetal05} and obtained regular
\Swift\ monitoring.
\psr\ is serendipitously and intentionally inside the \Swift\ field of view
in many observations to date (see Table~\ref{tab:obsx}).
These observations have exposure times ranging from 1~ks to 11~ks, and
the source is detected in them.
In each case, the 0.3--10~keV count-rate is determined using aperture
photometry.
A source region of radius 8~pixels centred on the position of \psr\ is
used and compared to a nearby background region of radius 63~pixels.
Derived count-rates are listed in Table~\ref{tab:obsx}.

\begin{figure}
 \includegraphics[width=\columnwidth]{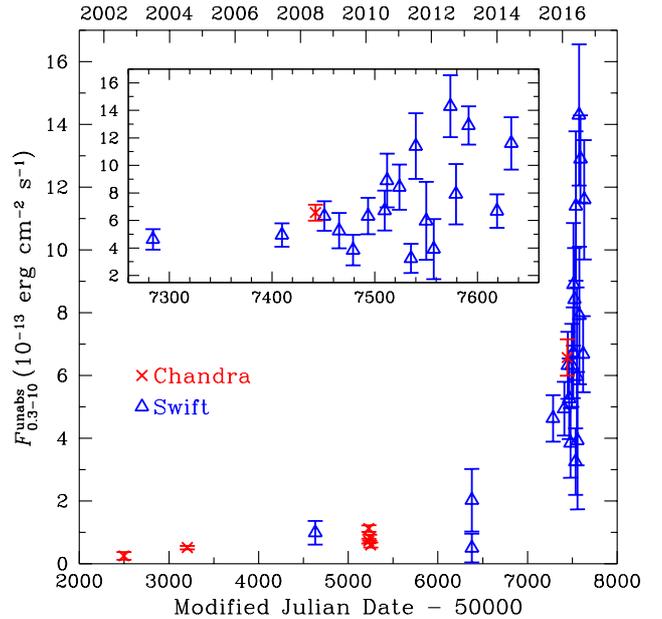}
 \caption{
X-ray light curve of \psr/\mttwo\ from 2002 to 2016.
Points (and $1\sigma$ error bars) are \Chandra\ (crosses) and \Swift\
(triangles) unabsorbed 0.3--10~keV flux (see Table~\ref{tab:obsx}).
Inset: Closer view of the data covering the period from 2015 September
to 2016 September.
}
 \label{fig:j2032lc}
\end{figure}

We use WebPIMMS to calculate the unabsorbed 0.3--10~keV flux for each
\Chandra\ and \Swift\ count-rate, assuming
$N_{\rm H}=7.7\times 10^{21}\mbox{ cm$^{-2}$}$ and $\Gamma=2$ power law
(see Table~\ref{tab:spectrafit}),
since the total number of counts in each observation is usually $<40$.
The resulting values are listed in Table~\ref{tab:obsx} and shown in
Fig.~\ref{fig:j2032lc},
and we see that the current X-ray flux is a factor of $\sim 14/0.2=70$ times
higher than in 2002 and $\sim 14/0.8\approx 20$ times higher than in 2010.
Because of the spatial resolution of \Swift, there are contributions to the
\Swift\ X-ray flux from the nearby sources shown in Fig.~\ref{fig:field}.
However this contamination is only significant in the 2008 and possibly
2013 data.
As we see from Fig.~\ref{fig:nearx}, the brightest sources in the field
other than \psr\ prior to 2015--2016 are Cygnus OB2~4 and MT91~221, and these
likely contribute to at most $\sim 9/(9+26)=25\%$ of the \Swift\
flux in the current epoch when \psr\ has clearly brightened.
We also checked that the recent flux decrease from mid to late May 2016 is not
reflected in a bright nearby source during the same period.

\section{H$\alpha$ equivalent width and disc size} \label{sec:optical}

\begin{figure}
\begin{center}
 \includegraphics[width=\columnwidth,angle=-90]{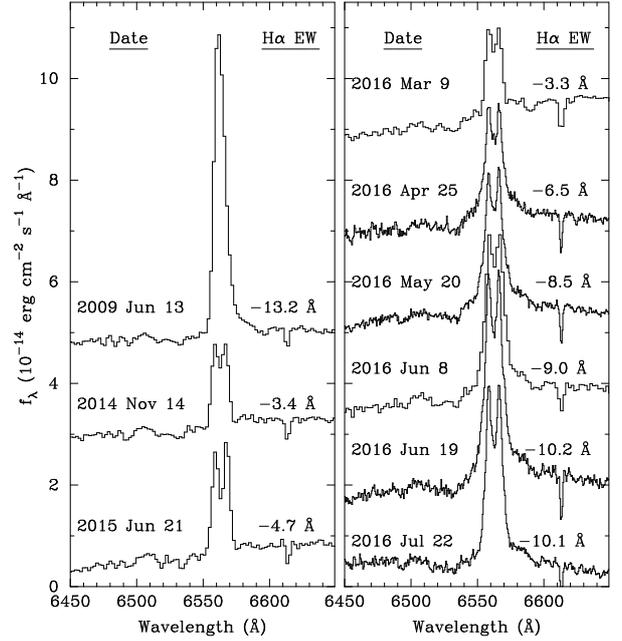}
\end{center}
 \caption{
H$\alpha$ region of spectra of \mttwo\ (see Table~\ref{tab:halpha}).
Spectra have been shifted vertically for clarity.
2009 spectrum is the same as appears in \citet{camiloetal09}.
Absolute flux densities for MDM spectra are not reliable due to the
narrow (1\arcsec) slit width used.
}
 \label{fig:halpha}
\end{figure}

Optical spectra of \mttwo\ were obtained on the MDM Observatory's
2.4~m and 1.3~m telescopes, as well as the Liverpool 2~m telescope,
with increased monitoring cadence in 2016.
Figure~\ref{fig:halpha} and Table~\ref{tab:halpha} show the H$\alpha$
emission line and \EW\ measurements.
The 2009 spectrum was published in \citet{camiloetal09}.
Here, we remeasured \EW\ from all spectra,
including the broad wings that become most apparent in 2016.
Evidently the circumstellar disc was largest in 2009, as indicated by the
stronger, single-peaked emission line.  The double-peaked structure of more
recent, weaker line profiles indicates a smaller disc, with peak separation
corresponding to $2v\,{\rm sin}\,i\approx360$~km~s$^{-1}$, where $v$ is
velocity  and $i$ is inclination angle.
The He~I~$\lambda5876$ is also double peaked in these spectra.

\begin{table*}
 \caption{
Optical spectra of \mttwo.
}
 \label{tab:halpha}
 \begin{tabular}{lllccc}
  \hline
Date & Telescope & Instrument & Resolution & H$\alpha$ \EW & $R_\mathrm{disc}/R_\mathrm{OB}$ \\
 & & & \AA & \AA & \\
\hline
2009 Jun 13 & MDM 2.4~m & Modspec & 3.5 & $-13.2$ & 13.1 \\
2014 Nov 14 & MDM 2.4~m & Modspec & 3.5 & $-3.4$ & 5.5 \\
2015 Jun 21 & MDM 2.4~m & Modspec & 3.5 & $-4.7$ & 6.8 \\
2016 Mar  9 & MDM 1.3~m & Modspec & 3.5 & $-3.3$ & 5.4 \\
2016 Apr 25 & Liverpool 2~m & FRODOspec & 1.2 & $-6.5$ & 8.3 \\
2016 May 20 & Liverpool 2~m & FRODOspec & 1.2 & $-8.5$ & 9.9 \\
2016 Jun  8 & MDM 2.4~m & Modspec & 3.5 & $-9.0$ & 10 \\
2016 Jun 19 & Liverpool 2~m & FRODOspec & 1.2 & $-10.2$ & 11 \\
2016 Jul 22 & Liverpool 2~m & FRODOspec & 1.2 & $-10.1$ & 11 \\
  \hline
 \end{tabular}
\end{table*}

\mttwo\ does not always have detectable emission lines.  They were not
reported in \citet{masseythompson91}, although that study did not cover
the H$\alpha$ region.  Spectra at additional epochs were reported by
\citet{salasetal13} under a different name (Cygnus OB2-4~B) for the star:
In 2012 September it showed no emission lines, while in 2008 October and
2013 September--October, double-peaked H$\alpha$ and H$\beta$ were present.

\begin{figure}
 \includegraphics[width=\columnwidth]{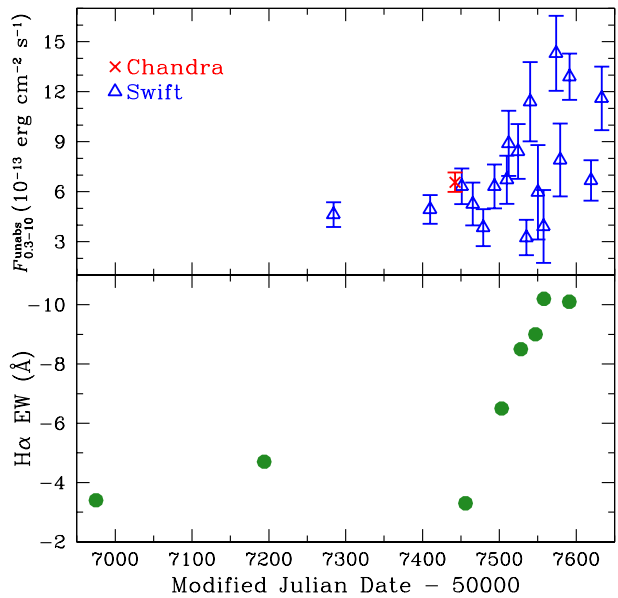}
 \caption{
X-ray and optical light curves of \psr/\mttwo\ from late 2014 to 2016.
Top: Points (and $1\sigma$ error bars) are \Chandra\ (crosses) and \Swift\
(triangles) unabsorbed 0.3--10~keV flux (see Table~\ref{tab:obsx}).
Bottom: Circles are H$\alpha$~\EW\ measurements (see Table~\ref{tab:halpha}).
}
 \label{fig:xoptlc}
\end{figure}

We estimate the size of the circumstellar disc around \mttwo\ using the
relation from \citet{hanuschik89}, which connects circumstellar disc radius
$R_\mathrm{disc}$ and stellar radius $R_\mathrm{OB}$ to the H$\alpha$
\EW, i.e.
\begin{equation}
\log\left(R_\mathrm{disc}/R_\mathrm{OB}\right) = 0.4+0.64\log\left(-\EW\right).
\end{equation}
Our \EW\ measurements of \mttwo\ indicate that its circumstellar disc
has varied in size by more than a factor of two, from $5$ to $13R_\mathrm{OB}$
(see Table~\ref{tab:halpha}).
For typical B0~V stars, $R_\mathrm{OB}\sim 8\,R_{\sun}$ \citep{vaccaetal96},
which yields a range of $R_\mathrm{disc}$ from
$\sim 40\,R_{\sun}=3\times 10^{7}\mbox{ km}=0.2\mbox{ AU}$
to $\sim 100\,R_{\sun}=7\times 10^{7}\mbox{ km}=0.5\mbox{ AU}$.
Approximately similar values of $R_\mathrm{disc}$ are obtained using the
separation of the double peaks seen in spectra from 2014 to 2016.
However for these later spectra, \EW\ varies by a factor of about three
(implying disc size change of about two), while peak separation does not
change significantly.
It is also noteworthy that broad wings appear in the H$\alpha$ spectra,
beginning in 2016 Apr 25, when \EW\ increases (see Fig.~\ref{fig:xoptlc},
which shows X-ray and optical light curves since late 2014).
These broad wings could be due to higher wind velocities, which could
cause the observed increase in X-ray emission.

\section{Discussion} \label{sec:discuss}

In this work, we described recent multiwavelength observations of the
high-energy binary system containing the 143~ms radio pulsar \psr\ and
the B0~Ve companion star \mttwo.
The orbit is very eccentric ($0.94<\epsilon<0.99$) and large
($P_{\rm orb}\approx 45-50$~yr), and the pulsar is accelerating rapidly
towards periastron passage in 2017 November.
We updated orbital parameters of the system obtained via radio monitoring,
combined with \Fermi\ observations, of the pulsar.
Archival and recent \Chandra\ and \Swift\ observations show that the
\psr/\mttwo\ system has brightened significantly in X-rays, especially
within the last year as it approaches periastron.

We can understand the current and past behavior of the \psr/\mttwo\ system
by comparing it to the high-energy binary pulsar system \psrb/LS~2883,
since the two systems have many similarities.
\psrb\ is a 47.76~ms radio pulsar in an eccentric ($\epsilon\approx 0.87$)
3.4~yr orbit with LS~2883, which is a 09.5~Ve star
\citep{johnstonetal92,negueruelaetal11}.
\psrb/LS~2883 has been observed across the electromagnetic spectrum during
each periastron passage since its discovery
(see, e.g. \citealt{chernyakovaetal15}, for discussion of the most recent
passage in 2014), as well as around apastron
(see, e.g. \citealt{hirayamaetal99}).
The resulting studies show that its multiwavelength emission can be understood
as shock interaction between the relativistic wind emitted by the pulsar and
the circumstellar disc and wind of the companion star \citep{tavaniarons97}.
However, in attempting to extrapolate the observed behavior of \psrb\ to
that of \psr, it is important to note that the binary separation at
apastron is $\sim 11$~AU in the case of \psrb, while \psr\ has only been
observed up to this point at a binary separation $>10$~AU.
At this large distance, the wind from \mttwo\ is possibly tenuous,
and its collision with the pulsar wind is possibly weak.
Also LS~2883 is four times more luminous and a different stellar type
\citep{negueruelaetal11}, so its wind properties are likely different
from that of \mttwo.

For \Chandra\ observations taken in 2002, 2004, and
around 2010, \psr\ was quite distant from \mttwo, at a binary separation
of $\gtrsim 30$~AU (see Fig.~\ref{fig:orbit}).
Thus the observed X-ray emission
[with $L_{\rm X}\approx(0.05-0.2)\times 10^{32}\mbox{ erg s$^{-1}$}
(d/1.3\mbox{ kpc})^2$]
could be due to what is effectively an isolated Be star or an isolated
young pulsar.
In the former case, bright X-ray emission from powerful wind shocks of O
stars is quite common, and spectra are often fit with an optically thin
thermal plasma model (as performed in Section~\ref{sec:chandra});
the source of X-ray emission from stars of later stellar types is less
certain, with a transition around early B stars, like \mttwo, that have
$L_{\rm X}\sim 10^{30}-10^{32}\mbox{ erg s$^{-1}$}$
\citep{berghoferetal97,gagneetal11,nazeetal14}.
Using the relation between X-ray and bolometric luminosities,
$\log L_{\rm X}/L_{\rm bol}\approx -7.2$, found for O and bright B stars
(although there is large dispersion at the luminosity of stars similar to
\mttwo; \citealt{rauwetal15}),
we find $L_{\rm X}=4\times 10^{30}\mbox{ erg s$^{-1}$}$, which matches the
X-ray luminosity of \mttwo\ in 2002.

For isolated pulsars,
X-ray radiation can have non-thermal and thermal contributions.
Non-thermal emission can be generated by a relativistic wind,
which produces a ratio between X-ray luminosity to rotational energy loss
of $L_{\rm X}/\dot{E}\lesssim 10^{-3}$ and a spectrum that is best-fit by
a power-law with $\Gamma\approx 1-3$ \citep{becker09}.
For \psr, $\dot{E}=1.5\times 10^{35}\mbox{ erg s$^{-1}$}$, and the measured
power law is $\Gamma\approx 1.5-2.5$ (Table~\ref{tab:spectrafit}).
Thus a pulsar wind can easily be the source of observed X-rays.
Meanwhile, thermal emission
for intermediate age ($\sim 10^5$~yr) neutron stars gives
$L_{\rm X}\sim 10^{31}-10^{33}\mbox{ erg s$^{-1}$}$
(see, e.g. \citealt{potekhinetal15}).

The more recent observations since late 2015 show significant brightening
in X-rays (see Fig.~\ref{fig:j2032lc}), with
$L_{\rm X}\approx(0.6-2.9)\times10^{32}\mbox{ erg s$^{-1}$}
(d/1.3\mbox{ kpc})^2$.
Although these luminosities are somewhat lower than the X-ray luminosity seen
for \psrb\ at apastron
($L_{\rm X}\approx 5\times 10^{32}\mbox{ erg s$^{-1}$}$;
\citealt{hirayamaetal99,uchiyamaetal09}),
the brightening of \psr\ suggests that the pulsar has entered the regime
where the pulsar wind is interacting strongly with the Be star wind.
X-ray spectral studies of \psrb\ find that most observations made by
\Chandra, \textit{Suzaku}, \Swift, and \textit{XMM-Newton} can be fit
with a power law model
(see, e.g. \citealt{tavaniarons97,chernyakovaetal09,chernyakovaetal14}).
\citet{chernyakovaetal06,chernyakovaetal09} and \citet{uchiyamaetal09}
show that the value of $\Gamma$ varies with orbital phase and that
$\Gamma$ changes from 1.8 around apastron to 1.2 right before the pulsar
enters the circumstellar disc of LS~2883 at periastron,
a decline that is similar to what is found from our limited spectra of
\psr\ (see Table~\ref{tab:spectrafit}).

The \psr/\mttwo\ system will continue to be monitored across the
electromagnetic spectrum as the pulsar approaches periastron.
When it is near periastron, the system may brighten even more if
the pulsar accretes from the circumstellar disc of the Be star.
\psr/\mttwo\ will thus serve as an invaluable tool for comparing and
contrasting to the
very well-studied and previously unique gamma-ray binary \psrb/LS~2883.

\section*{Acknowledgements}
The authors are indebted to Paul Ray for his support and careful reading
of the manuscript.
The authors thank Maria Chernyakova for helpful comments
and Belinda Wilkes and the \Chandra\ team for conducting the
2016 DDT observation.
WCGH acknowledges support from the United Kingdom Science and Technology
Facilities Council (UK STFC).
Pulsar research at Jodrell Bank Centre for Astrophysics is supported by a
Consolidated Grant from UK STFC.
The \Fermi\ Collaboration acknowledges generous ongoing support from a
number of agencies and institutes that supported both development and operation
of the LAT as well as scientific data analysis.
These include the National Aeronautics and Space Administration and
Department of Energy in the United States,
Commissariat \`{a} l'Energie Atomique and Centre National de la Recherche
Scientifique/Institut National de Physique Nucl\'{e}aire et de Physique
des Particules in France,
Agenzia Spaziale Italiana and Istituto Nazionale di Fisica Nucleare in Italy,
Ministry of Education, Culture, Sports, Science and Technology (MEXT),
High Energy Accelerator Research Organization (KEK), and
Japan Aerospace Exploration Agency (JAXA) in Japan, and
K.~A.~Wallenberg Foundation, Swedish Research Council, and
Swedish National Space Board in Sweden.
Additional support for science analysis during the operations phase is
gratefully acknowledged from the Istituto Nazionale di Astrofisica in Italy
and Centre National d'\'Etudes Spatiales in France.
The Liverpool Telescope is operated on the island of La Palma by Liverpool
John Moores University in the Spanish Observatorio del Roque de los Muchachos
of the Instituto de Astrofisica de Canarias with financial support from UK STFC.

\label{lastpage}
\end{document}